\documentclass[a4paper,conference]{IEEEtran}

\usepackage{graphicx}
\usepackage{subfig}
\usepackage{mathtools}

\usepackage{color}
\usepackage{algorithmicx}
\usepackage[ruled]{algorithm}
\usepackage{algpseudocode}
\algnewcommand\algorithmicinput{\textbf{Input:}}
\algnewcommand\INPUT{\item[\algorithmicinput]}

\begin{document}

\title{\ Prediction Defaults for Networked-guarantee Loans}
\author{\IEEEauthorblockN{Dawei Cheng\IEEEauthorrefmark{1},
Zhibin Niu\IEEEauthorrefmark{2},
Yi Tu\IEEEauthorrefmark{1} and
Liqing Zhang\IEEEauthorrefmark{1}}
\IEEEauthorblockA{\IEEEauthorrefmark{1}
Key Laboratory of Shanghai Education Commission for Intelligent Interaction and Cognitive Engineering,
\\Department of Computer Science and Engineering, Shanghai Jiao Tong University, China.
\\Email: \{dawei.cheng, tuyi1991, lqzhang\}@sjtu.edu.cn }
\IEEEauthorblockA{\IEEEauthorrefmark{2} School of Computer Software, Tianjin University, China.
Email: zniu@tju.edu.cn}
}

\maketitle
\begin{abstract}
Networked-guarantee loans may cause the systemic risk related concern of the government and banks in China. The prediction of default of enterprise loans is a typical extremely imbalanced prediction problem, and the networked-guarantee make this problem more difficult to solve. Since the guaranteed loan is a debt obligation promise, if one enterprise in the guarantee network falls into a financial crisis, the debt risk may spread like a virus across the guarantee network, even lead to a systemic financial crisis. In this paper, we propose an imbalanced network risk diffusion model to forecast the enterprise default risk in a short future. Positive weighted k-nearest neighbors (p-wkNN) algorithm is developed for the stand-alone case -- when there is no default contagious; then a data-driven default diffusion model is integrated to further improve the prediction accuracy. We perform the empirical study on a real-world three-years loan record from a major commercial bank. The results show that our proposed method outperforms conventional credit risk methods in terms of AUC. In summary, our quantitative risk evaluation model shows promising prediction performance on real-world data, which could be useful to both regulators and stakeholders.
\end{abstract}
\IEEEpeerreviewmaketitle

\section{Introduction.}
One of the most important drivers of macroeconomic conditions and systemic risk is consumer spending, which accounted for over two thirds of American gross domestic product\cite{khandani2010consumer}. The opportunities and risk exposures in consumer lending are equally outsized, which makes it necessary to rely on models and algorithms rather than human discretion. Consumer credit risk evaluation is often technically addressed in a data-driven fashion and has been extensively investigated \cite{baesens2003using, hand1997statistical}.

As one of major part in consumer lending business, the economic and banking importance of the small and medium enterprise (SME) loans is well recognized in academic and policy literature \cite{biggs2002small, ruzzier2006sme}. However, the existing mechanism for loan decision-making falls behind business demand. Most of the criteria are designed for major independent players, while the less formal small and medium enterprises may provide inaccurate or manipulated information.

\begin{figure}[tb!]\vspace{-10pt}
  \centering
  \includegraphics[width=0.9\linewidth]{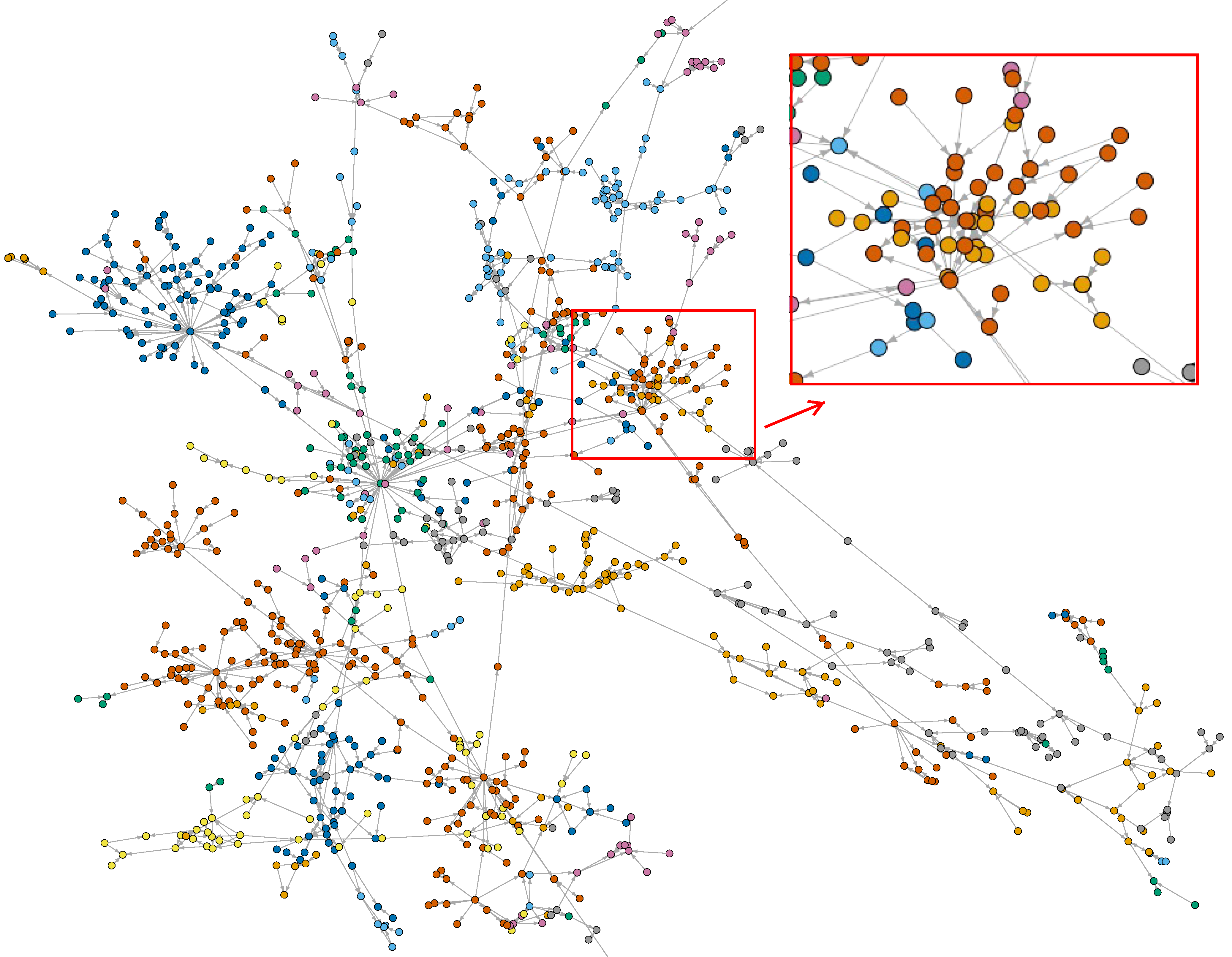}
  \caption{A real-world loan guarantee network formed from bank
records, with each node as an enterprise. Zoom in for more details.}\label{introduction-networks}\vspace{-15pt}
\end{figure}

In order to meet the criteria of financial security by the banks, groups of small and medium enterprises back each other. When more and more enterprises are involved, they form networks with complex structures. Figure \ref{introduction-networks} gives a guarantee network with more than 600 enterprises bound together in a connected network. Such a complex network is a double-edged sword to the national economy. On the one hand, these secured loans can help an enterprise to source financing rapidly and promote development during a period of economic growth. On the other hand, although a complex network can slow down the risk of corporate default during a period of economic downturn, it may also lead to large-scale defaults and spread the infection.

Abundant machine learning approaches have been studied for credit risk on SMEs \cite{deyoung2015risk, dcheng2018pvis, wu2014business}. As features in financial aspects are not as high as tasks in image and speech domains, learning algorithms with high interpretability and low complexity is required \cite{kovalerchuk2000data}. The K-Nearest Neighbor(KNN) classifier is one of the most fundamental and simple classification methods and should be one of the first choices for a classification study when there is little or no prior knowledge about the distribution of the data \cite{peterson2009k}. It is one of the popularest in financial domain because of the high performance on low dimensional data, simple to implement and explanatory, compared with logistic regression, SVM, random forest and neural network.

However, all above approaches are designed with the assumption that the number of samples of each class is roughly the same. This is not the case in financial scenarios, loan default records is much smaller than normal repayments. Our statistics show that only about 6\% of borrowers have default records. Such highly skewed datasets need tailored algorithms for data mining. Sampling-based approach is widely used in industry and academic research \cite{laurikkala2001improving, han2005borderline, he2009learning}. But it inevitably change the distribution of the overall dataset. Other approaches including cost sensitive learning \cite{zhou2006training} and ensemble learning \cite{yan2011incremental} are suffer from high model complexity and low interpretability.

Besides, there are some studies on networked guarantee\cite{levitsky1997credit, niu2017hybrid}, but most of them focus on statistical analysis \cite{meng2015credit} or revel some structure feature of the network \cite{garcia2010credit}. Little of them research on predicting credit risk as well as the risk diffusion over the network. In the social network literature, some studies address the problem of information diffsion over social network \cite{tang2009social, irfan2011game, xiao2016modeling}, but as a financial behavior, risk diffusion from economic network is greatly different from social network.

In this paper, in order to tackle above challenges, we propose INDDP (Imbalance Network Diffusion Default Prediction). Rich financial features, including network related measures, are first feed to P-wKNN(positive weighted k-Nearest-Neighbor Classification and Regression method), which is easy to implement and naturally designed for imbalanced datasets without biased sampling. Then, a probabilistic graph model is proposed to model the risk diffusion over networks.

In a nutshell, the main contributions of this paper are:

\begin{enumerate}
  \item To the best of our knowledge, this is the first study to use probabilistic graph model to predict credit risks diffusion on guarantee network.
  \item We propose a novel positive weighted k-Nearest-Neighbor Classification and Regression method for probability estimation, which is easy to implement, naturally designed for imbalanced data and with high interpretability.
  \item We propose a practical solution to evaluate default risks for networked-guarantee loans. which is much simpler than the classic physical diffusion model and more effective for predicting defaults in networked-guarantee loan data.
\end{enumerate}

\section{Preliminaries.}

We briefly introduce the related financial business procedure and our feature description.

\subsection{Business procedure.}

In order to obtain a loan, a borrower needs to open an account and provide detailed information to the bank. Banks may be reluctant to issue the loan, as it is difficult for SMEs to meet the bank’s lending criteria, which are designed for big companies. There is something of a blank area for setting the criteria for SMEs due to their lack of security. However, they are permitted to offer other corporations as endorsement. Usually, banks need to collect as much fine-grained information as possible, including transaction information, customer information, asset information such as mortgage status, and history of loan approval.

\begin{figure}[tb!]\vspace{0pt}
  \centering
  \includegraphics[width=1\linewidth]{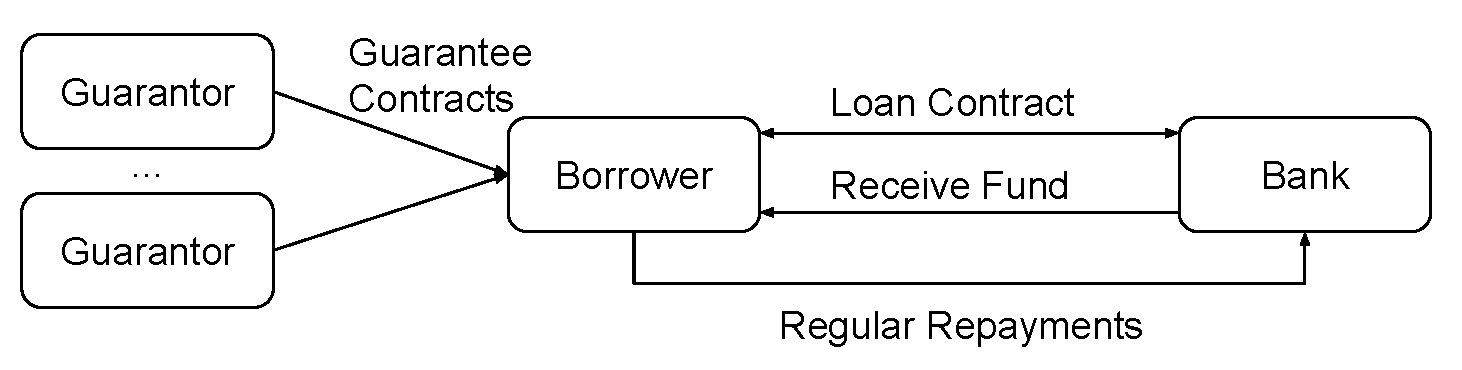}
  \caption{Guarantee loan process. The SME (borrower) wishing to get a bank loan first needs to sign guarantee loan contracts with guarantors before signing loan contracts. The company will repay the loan in installments.}\label{preliminaries-procedure}\vspace{-15pt}
\end{figure}

As Figure~\ref{preliminaries-procedure} shows, there is often more than one guarantor per loan transaction, and a single guarantor may make several loan transactions in a period. Upon approval of the loan, the SMEs usually obtain the full loan amount immediately, and start making repayments by regular installments until the end of the loan contract. We reconstruct the guaranteed network based on nine data tables: customer profile, loan account information, repayment status, guarantee profile, customer credit, loan contract, guarantee relationship, guarantee contract and default status.

\subsection{Data Description.}

We collect loan records spanning three years from a major commercial bank in China. The names of the customers in the records are encrypted and replaced by an ID; we can access basic profile information such as the enterprise scale, and loan information such as the guarantee ID and the loan credit.

In order to build a feature that can reliably represent the statistical relationships between the customer information and their repayment ability, we cleaned the data and constructed the features including the following factors:

 \textbf{Registration information:} This includes the character, capital, collateral, capability, condition and stability. We use business nature, registered capital, enterprise scale, employee number and other features as a corporation’s basic profile. Most banks require the company to update this basic information when the enterprise makes a loan application; we use the latest information.

\textbf{Historical behavior:} This includes credit history, default records, default amount, total loan amount, loan count, total loan frequency, and total default rates. They are calculated with all the loan records before the current one.

\textbf{Network related measures:} For example, the centralities are extracted. As discussed above, the basic profile may not be completely trustworthy, as the small business might provide out of date or even fake information to the bank. However, the guaranteed network offers trustworthy information, as the bank can build this from its own records.

\section{Algorithm.}
As loan and repayment behaviors are temporal events, we collect all these events in the current time windows as training datasets, then predict the probability of defaults in the next time window. We define a default event as the record that the borrower did not refund the repayment in time according to loan contracts. We address the following problems:
\\
\textbf{Given:} a subsequence of event stream $E_{-\omega}$ , where $\omega$ is a user-defined time window.
\\
\textbf{Predict:} the probability $e^{*}$  of a default occurrence in the next time window $\omega$.

Predicting default can be defined as estimating the conditional probability $P(\exists e^*\in E_{+\omega}|E_{-\omega})$. Due to the unbalanced data, there is insufficient data to estimate an accurate conditional distribution. We first model the conditional distribution by our tailored nearest-neighbor-based algorithm. Then, we model the diffusion distribution using a generative model.
\vspace{-5pt}
\subsection{Default modeling.}

As explained before, a loan guarantee is a promise by the guarantor to assume the debt obligation if the borrower defaults, and the default may spread across the network in the manner of a viral infection. The probability that the business will default (not repay the bank in time) is noted as $P(A)$, where $A$ is a small business.
\begin{equation}\label{e3.1}
  P(A) = P_s(A)+(1-P_s(A))\cdot P_g(A)
\end{equation}
Where $P_s(A)$ is its static default possibility, which means the default probability is caused by the self state of the business. $P_g(A)$ is the diffusion probability of how likely $A$ will be influenced when the guarantee of $A$ has defaulted. As $A$ provides a guarantee for other borrowers, $A$ is responsible for repaying the debt if the borrowers default; this may also lead to $A$ defaulting. $P_g(A)$ is designed to represent this type of default probability. $P_g(A|x)$ denotes the conditional probability when a specific guarantee $x$ defaults. $P_g(A)$  can be calculated by estimating $P_g(A|x)$ and $P_s(x)$ for all of its warrantees.

\vspace{-10pt}
\begin{equation}\label{e3.2}
  P_g(A) = 1- \prod_{i=1}^{m}{(1-P_s(x_i)P_g(A|x_i))} \vspace{-5pt}
\end{equation}
Then, we bring function $g(x)=1-x$ for simplified representation and define the $d-th$  order $i-th$  neighbors of $A$ in out-degree as $N_{A,d,i}$. Therefore, all neighbors of $A$ whose distances are smaller than $d-th$ order are notated as $\textrm{O}_d(A)=\{x_i|N_{A,d,i}, i \in 1:m\}$. Equation \ref{e3.2} can be reformulated as:

\vspace{-10pt}
\begin{equation}\label{e3.3}
  P_g(A) = g\left(\prod_{x_i\in \textrm{O}_d(A)}{g\left(P_s(x_i)P_g(A|x_i)\right)}\right) \vspace{-5pt}
\end{equation}

Based on the equation \ref{e3.1} and \ref{e3.3}, we approximate the desired conditional distribution with the following joint distribution. The joint distribution of $A$ is:

\vspace{-10pt}
\begin{equation}\label{e3.4}
\begin{aligned}
  P(A) & = P_s(A) + \\
    & g(P_s(A))\cdot g\left( \prod_{x_i\in \textrm{O}_d(A)}{g\left(P_s(x_i)P_g(A|x_i)\right)}\right)
\end{aligned}
\end{equation}

\subsection{Sliding window scheme.}

Prediction shall be adapted to a dynamic setting with regularly updated forecasting results. Using a sliding window is a typical way for rolling prediction and is commonly adopted in event prediction practices. The business often runs on a quarterly basis; thus it would be helpful to know which borrowers might default on a quarterly basis from a business demand perspective.

\begin{figure}[tb!]\vspace{0pt}
  \centering
  \includegraphics[width=1\linewidth]{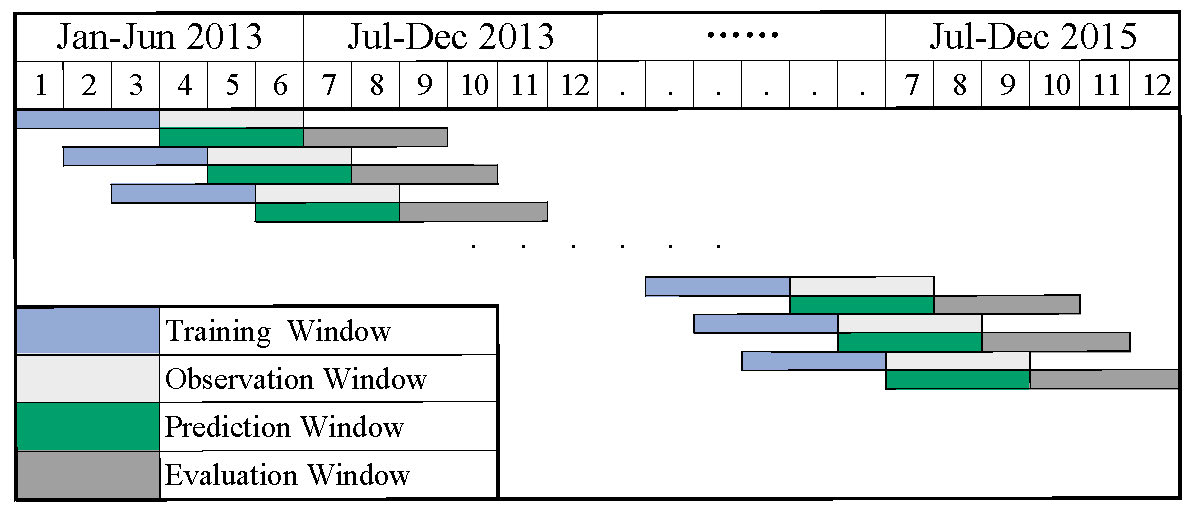}
  \caption{Illustration for the rolling sliding windows protocol. Features are extracted in the training window; the corresponding outcome default label is collected in the observation window. The features and default outcome are used to train the model. The trained model is used by collecting the input features during the prediction window and verifying their performance when we reach the end of the evaluation window.}\label{algorithm-timewindow}
\end{figure}

In this work, we set a time window of $\omega$ days. Features are extracted in the training window; the corresponding outcome default label is collected in the observation window. The features and default outcome are used to train the model. The trained model is used by collecting the input features during the prediction window and verifying their performance when we reach the end of the evaluation window.

Specifically, in this paper we use a three-month window for training, observation, prediction, and evaluation. As Figure~\ref{algorithm-timewindow} shows, in the training stage, for all customers who obtain bank loans from 2013 Q1 (first quarter of 2013), the features are extracted in that period, and the repayment status between 2013 Q2 are the labels to train the model. In the testing stage, we use the trained model to predict the customers who obtain loans between 2013 Q2 and use the real repayment status from 2013 Q3 to evaluate the performance when reaching the end of September 2013.

There are two reasons for such a sliding window setting:

\vspace{-2pt}
\begin{enumerate}
  \item Prediction shall be adapted to a dynamic setting with regularly updated forecasting results. In fact, using a sliding window is a typical way for rolling prediction and is commonly adopted in event prediction practices such as \cite{yan2015sales}.\vspace{-2pt}
  \item After observing from our datasets and acquiring feedback from financial experts, the business often runs on a quarterly basis. Thus, from a business demand perspective, it would be helpful to know which borrowers may default on a quarterly basis.\vspace{-5pt}
\end{enumerate}

\subsection{Estimating parameters.}

In order to obtain $P(A)$, the main task is to estimate $P_s(A)$ and $P_g(A|x)$ as present in Equation \ref{e3.4}.

\textbf{Estimating $P_s(A)$:} Estimating this probability requires predicting how the current event sequence will evolve over the next ω days, as this influences whether defaults occur or not. Given the current subsequence $E_{-\omega}$, we designed a positive weighted k-nearest-neighbor classification and regression (P-wkNN) approach.

\begin{figure}[tb!]
  \centering
  \includegraphics[width=1\linewidth]{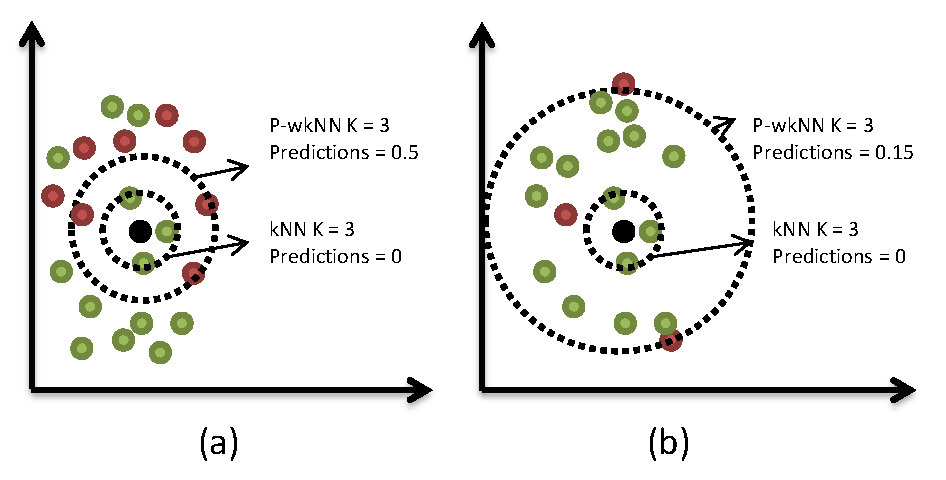}\vspace{-10pt}
  \caption{Illustration of the P-wkNN algorithm.} \label{algorithm-pwknn}\vspace{-20pt}
\end{figure}

Naive kNN, the non-parametric lazy learning algorithm, updates a target sample label by its top k nearest neighbors' label. For extremely imbalanced sample cases, negative samples (majority class) are much more frequent than positive samples (minority class). How to set the hyper parameter value of k is a tradeoff between performance and efficiency, for example, if k is too small, as the negative samples may overwhelm its near positive samples; if k is too large, the computation may be too complex. Hence, if we get top k positive samples, we could update a precise value for the sample. Figure 3 illustrates the algorithm if the target value is more likely to be a positive sample. P-wkNN could predict its value to 0.5 when k is set to 3 and weight set to equal, while kNN predicts 0. In situation b, the predicted instances are more likely to be a negative sample. If we set k to 3 and weights to equal, p-wkNN predicts 0.15 and kNN predicts 0. P-wkNN would not be influenced by outliers.

The detail of the algorithm is described below:

\begin{enumerate}
  \item Let $S=\{(x_i,y_i), i=1,2,\ldots,n\}$ be the training set and $x_i$ be the observations with a class label $y_i$. Let $x$ be a new observation; we need to predict its label $y$.
  \item Find the $k$ nearest positive neighbors to $x$ by a distance function $d(x,x_i)$ .
  \item Choose the $k-th$ positive neighbor’s distance $\sigma=d(x,x_{dk})$  as the threshold. So the defined neighbors of $x$ are $N=\{x_i|d(x,x_i)\leq\sigma, i=1,2,\ldots,n\}$.
  \item Normalize the distance by function $D_i=D(x,x_i)=\frac{d(x,x_i)}{\sigma} \in [0,1]$. Then transform $D_i$ with a kernel function $K(\cdot)$ into a weight $w_i=K(D_i)$ .
  \item Calculate the total weighted class label. Let   be the predicted class label; we have two methods to give its estimations. For the classification method, we have $y=\max _r\left(\sum_{x_i \in N}{w_i I(y+i=r)} \right)$  and for the regression method $y=\frac{\sum_{x_i \in N}{w_iy_i}}{\sum_{x_i \in N}{w_i}}$.
\end{enumerate}

\begin{figure}[tb!]
  \centering
  \includegraphics[width=1\linewidth]{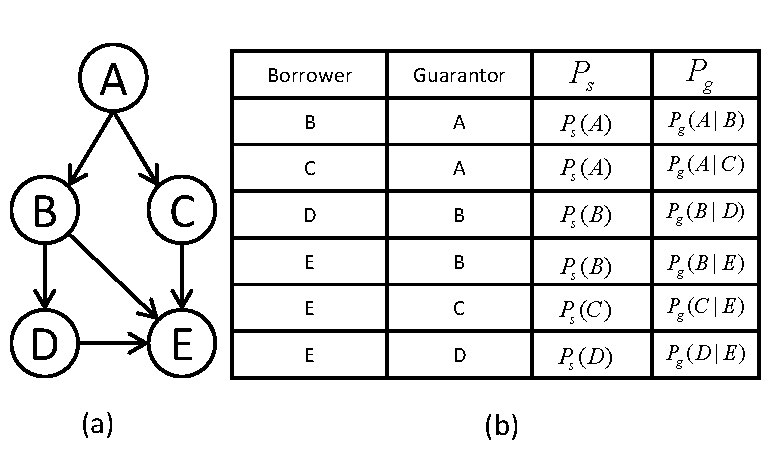}\vspace{-10pt}
  \caption{(a). a typical guarantee network. (b). Table $X$ of the loan guarantee network shown left.} \label{algorithm-orders}\vspace{-20pt}
\end{figure}

\textbf{Estimation $P_g(A|x)$:} the likelihood of defaults diffusion from $x$ to $A$. Once the warrantee has defaulted, a label column is added to the filtered records table $X$ as shown in Figure \ref{algorithm-orders} (b). The label’s value is decided by its guarantor's situation in a given time window $\omega$. If the guarantor defaults, the label of this warrantee record will be set to 1, otherwise 0.

We extract the features of the edges from the guarantors to the warrantees, including guaranteed amount, loan amount, warrantees’ degrees, guarantors’ degrees and essential company registration information from both sides. We apply P-wkNN to train the diffusion probability. For a new guarantee relationship, we find its k nearest positive neighborhoods, then estimate the value as described in P-wkNN algorithm.

\subsection{Make predictions.}
In practice, a borrower may have one or more guarantor, making a set of directed acyclic graphs, an example of which is shown in Figure \ref{algorithm-orders} (a). The cyclic graph in a guarantee network is forbidden by administration of regulation departments. Default risk probability diffuses directly from the borrower to the guarantor, which means the probability update needs to follow the signal directed dependency chain. As shown in Figure \ref{algorithm-orders}, node B can be both a borrower and a guarantor. If we want to obtain the default probability of node B, we need to calculate node E and node D in advance. Node D is dependent on node E at the same time. All this makes the update path much more complex.

\vspace{-0.1cm}%
\alglanguage{pseudocode}
\begin{algorithm}[h]
\caption{d-order diffusion approximation}
\label{Algorithm:D}
\begin{algorithmic}[1]
\INPUT $P_s(A), P_g(A|x)$ and order $d$.
\State $\textrm{\textbf{initilize:}}$
\State $P(A) \gets 0$
\State Table $X \gets \left(x,A,P_s(A),P_g(A|x)\right)$
\Procedure{FUN}{$A,d$}
  \If{$d=0$ or $A_{\textrm{outdegree}=0}$}
    \State\Return{$P_s(A)$}
  \ElsIf {$\sum_{i \in 1:m}{P(A|x_i)=0}$}
    \State\Return{$P_s(A)$}
  \Else
    \For{ \textbf{each} $x$ in $\textrm{O}_d(A)$}
        \State\Return{$\textrm{FUN}\left(x,d-1\right)$}
    \EndFor
  \EndIf
\EndProcedure
\Statex
\end{algorithmic}
\label{pseudocode}
  \vspace{-0.4cm}%
\end{algorithm}
\vspace{-10pt}

Therefore, we designed a d-order risk diffusion algorithm to predict default risks. This is a BFS (breadth-first-search) based recursion algorithm. First, given historical event streams, we estimate $P_s(A)$ and $P_g(A|x)$ as described in section 4.3, setting the hyper-parameter $d$, the diffusion order of networked-guarantee loans. Second, we construct table $X$ as the format of $\left(x,A,P_s(A),P_g(A|x)\right)$ and initialize the output probability $P(A)$ to 0. Third, we set the recursion stop conditions and main procedure. If diffusion order $d$ is equal to 0 or $A$ does not provide any guarantees for others, we return $P(A)$ by $P_s(A)$. Meanwhile, if all the $P_g(A|x_i)$ is 0, $P(A)$ is also equal to $P_s(A)$. In other conditions, for each node $x \in \textrm{O}_d(A)=\{x_i|N_{A,d,i}, i \in 1:m\}$, calculate $P(x)$ as given $P_s(x)$, $P_g(x|N_{x,d-1,i})$, which is estimated by section 4.3. For a simple illustration, we reformulate equation \ref{e3.4} as function $\textrm{FUN}(A)$, then we obtain the formulation below:
\begin{equation}\label{e4.1}
\begin{aligned}
  P(A) &= \textrm{FUN}(A) = \\
  & \textrm{FUN}\left(\textrm{FUN}(x|x \in \textrm{O}_d(A))\right)
\end{aligned}
\end{equation}
The pseudocode is given in algorithm~\ref{pseudocode}.

\subsection{Complexity analysis.}
Assuming there are total $n$ borrowers, each borrower $A$ has $m$ guarantors. As the number of recursions is different for each guarantor, we approximate each vertex out degree as $v_{out} $ and recursive depth as $Depth$. If the diffusion order $d$ is smaller than $Depth$, the function will be called $v_{out}^d$ times; otherwise, for each guarantor, the function will be called $v_{out}^{Depth}$. Therefore, the times of function call is the minimal of $n\cdot m\cdot v_{out}^{Depth}$ and $n\cdot m\cdot v_{out}^{d}$. Finally, the time complexity is $\textrm{O}\left(n\cdot v_{out}^{d}\right)$ and space complexity is $\textrm{O}(n)$ .

\section{Experiments.}

Empirical study is performed to evaluate the prediction performance. Next, we briefly review the dataset and then report the results.

\subsection{Data description}
The dataset covers the credit behavior of 23096 enterprises in the past three years. There are 3218428 bank loan records in total, 163665 of which have a default history, and 251743 guarantee records, with a total amount of 1147 billion dollars.

We predict default probability as explained earlier. For each small business, we use 3 sequences as a feature. We set the label to 1 if it defaults in the next 3 sequences and 0 otherwise. Then, we slide the window by 3 to generate features as predicting window as well as validation window for the next 3 sequences.

We measure the results in the validation window. All the small businesses with credit activities will be assigned a predicting probability of defaults. Then, we validate the probability with real repayment records and perform an AUC score for different approaches. Based on our experiment and domain knowledge, we set the number of nearest neighbor's $k$ from 1 to 10 and the diffusion degree $d$ from 0 to 6. The choice of $d$ is based on the analysis of the training data.

\begin{figure}[tb!]
  \centering
  \includegraphics[width=1\linewidth]{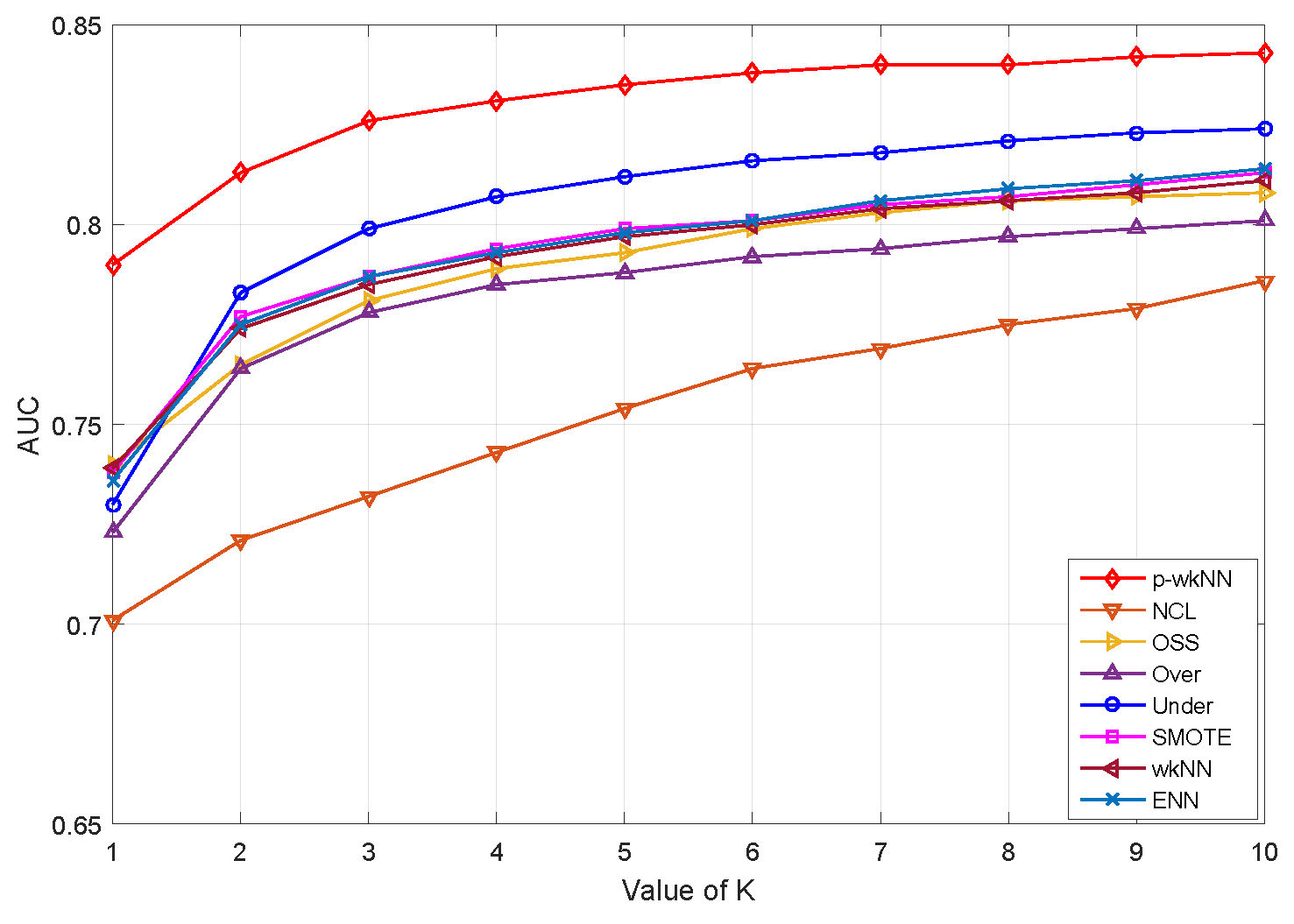}\vspace{-10pt}
  \caption{Compare with classic imbalanced learning approaches.} \label{experiments-pwknn}\vspace{-10pt}
\end{figure}

\subsection{Results.}
As mentioned earlier, the main challenges are the imbalanced data and the diffusion risk. We first compare our proposed P-wkNN with start-of-the-art imbalanced learning approaches, such as SMOTE \cite{chawla2003smoteboost}, over-sampling, under-sampling, One Sided Selection (OSS) \cite{kubat1997addressing}, Edited Nearest Neighbor (ENN) \cite{onnela2006complex} and Neighborhood Cleaning Rule (NCL) \cite{batista2004study}. Figure \ref{experiments-pwknn} gives the AUC curves. It is clear that our proposed pwkNN outperforms classic approaches with a much larger AUC value. In addition, we increase the value of $k$ to see its impact on performance. We can see that our approach always performs better than the other methods. All of them have a rapid lifting when the value of $k$ is smaller than 5, and the AUC gradually stabilizes when the $k$ becomes larger.

The second experiment is to explore the impact of diffusion. We alter the value of $k$ and d of our diffusion model and obtain the prediction results. We can see that a larger value of $k$ leads to better performance. Once the $k$ is greater than 5, the AUC becomes rather stable. For various $k$ values, taking the default diffusion into consideration ($d=1$) is much better than the assumption of no diffusion ($d=0$). This indicates that default diffusion does exist between the neighborhood borrower and the warrantee. However, the default does not diffuse across multiple nodes.

\begin{figure}[tb!]
  \centering
  \includegraphics[width=1\linewidth]{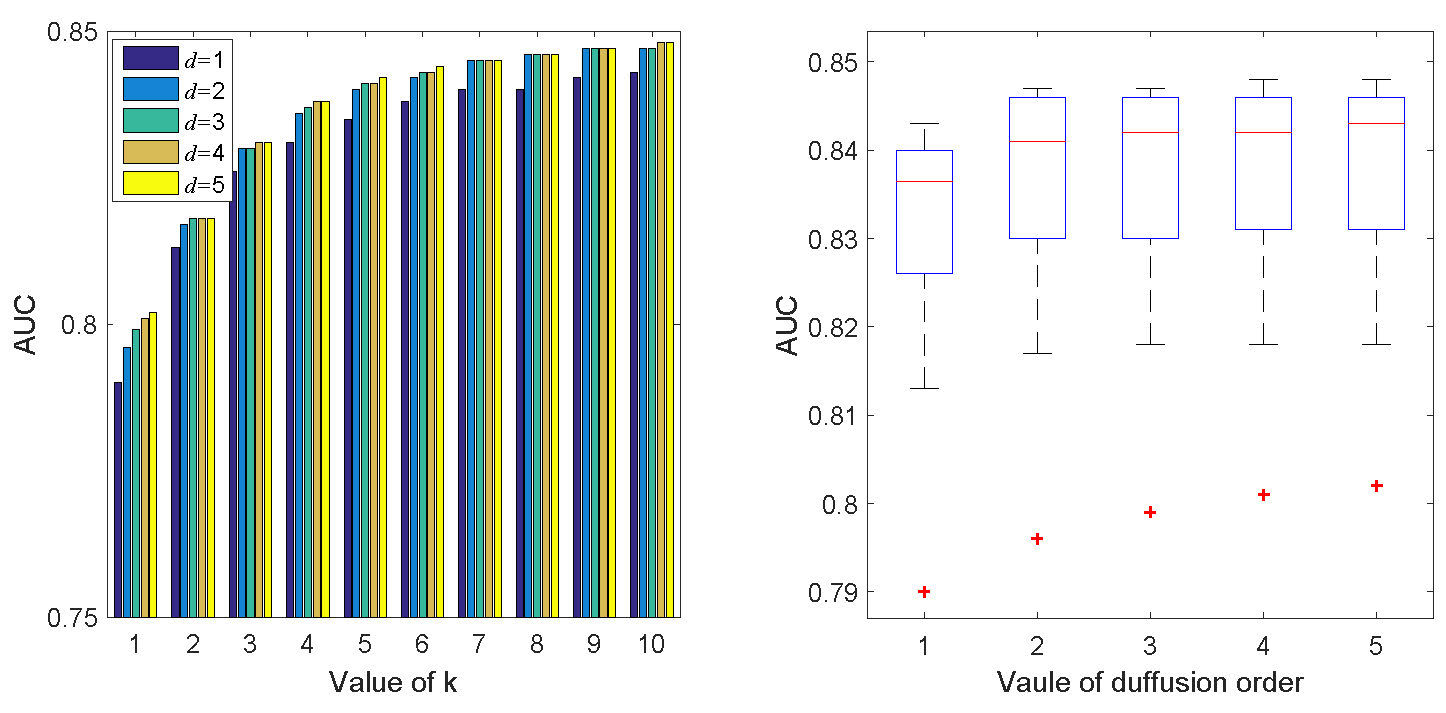}\vspace{5pt}
  \caption{Altering the diffusion order $d$ and value of pwkNN.} \label{experiments-order}\vspace{-10pt}
\end{figure}

Finally, we perform experiments to validate the effectiveness. We devise strong baseline algorithms and compare the precision, recall and F1 score with the following approach:

\begin{enumerate}
\item\textbf{INDDP}: Imbalance Network Diffusion Default Prediction, our proposed approach for networked-guarantee loan default prediction.
\item\textbf{P-wkNN}: our proposed classification model for imbalanced learning.
\item\textbf{wkNN}: Distance weighted k-NN algorithm, a refinement of the k-NN classification algorithm by weighting the contribution of each of the k neighbors according to their distance to the query point, giving greater weight to closer neighbors. Our P-wkNN is an extension of the original wkNN \cite{macleod1987re}.
\item\textbf{Random Forest (RF)}: an ensemble learning method for classification, regression and other tasks, which has proved effective in credit risk rating procedures \cite{jones2015empirical}.
\item\textbf{Support Vector Machines (SVM)}: a classical supervised learning model proven to be one of the most effective credit risk evaluation tools \cite{yu2011credit}.
\item\textbf{Logistic regression (LR)}: the classic credit evaluation model, widely used in the financial industry \cite{altman2007modelling}.
\end{enumerate}

We set $k$ to be 5 and $d$ to be 1 according to previous experiments. Our approach works best with such parameter setting. Table \ref{experiements-table1} summarizes the results: IDNNP outperforms all the baselines in terms of Precision, Recall and F1 score, indicating that considering the full scope of imbalanced and network diffusion is essential.

\begin{table} \vspace{6pt}
\centering
\caption{Results for default prediction.}\vspace{-5pt} \label{experiements-table1}
\begin{tabular}{l|c|c|c|c|c} \hline
\hline
 & Imbalance & Diffusion & Precision & Recall & F1-Score  \\ \hline
\hline
INDDP  & $\surd$ 	& $\surd$ 	 &  \textbf{0.81}  &  0.75   &  \textbf{0.78}  \\ \hline
P-wkNN & $\surd$ 	& -     	 &  0.72  &  0.75   &  0.73  \\ \hline
wkNN   & -      	& $\surd$ 	 &  0.64  &  \textbf{0.76}   &  0.69  \\ \hline
RF     & -  	    &   -    	 &  0.59  &  0.67   &  0.63  \\ \hline
SVM    & -  	    &   -    	 &  0.59  &  0.65   &  0.62  \\ \hline
LR     & -  	    &   -    	 &  0.54  &  0.63   &  0.58  \\
\hline
\hline\end{tabular}\vspace{-20pt}
\end{table}

\section{Conclusion.}
This paper presents a specifically tailored algorithm for automatically predicting default risks for networked-guarantee loans. As in many other financial tasks, the data is extremely imbalanced. The existing network and the potential default diffusion make risk assessment even more difficult. We adapt the classic KNN to handle the imbalanced datasets, using it to estimate network diffusion risks with a data-driven approach instead of forward modeling the diffusion physical model. We perform empirical experiments, give the best parameter settings for the networked-guarantee loan record, and compare it with state-of-the-art credit rating approaches. Future work will include refining the approach and experimenting for more sophisticated and imbalanced learning tasks.

\section{Acknowledgment}
The work was supported by the Key Basic Research Program of  Shanghai Science and Technology Commission, China (15JC1400103, 16JC1402800)  and the National Basic Research Program of China (2015CB856004).

\bibliographystyle{abbrv}
\bibliography{sigproc}

\begin{thebibliography}{10}

\bibitem{altman2007modelling}
E.~I. Altman and G.~Sabato.
\newblock Modelling credit risk for smes: Evidence from the us market.
\newblock {\em Abacus}, 43(3):332--357, 2007.

\bibitem{baesens2003using}
B.~Baesens, R.~Setiono, C.~Mues, and J.~Vanthienen.
\newblock Using neural network rule extraction and decision tables for
  credit-risk evaluation.
\newblock {\em Management science}, 49(3):312--329, 2003.

\bibitem{batista2004study}
G.~E. Batista, R.~C. Prati, and M.~C. Monard.
\newblock A study of the behavior of several methods for balancing machine
  learning training data.
\newblock {\em ACM Sigkdd Explorations Newsletter}, 6(1):20--29, 2004.

\bibitem{biggs2002small}
T.~Biggs.
\newblock Is small beautiful and worthy of subsidy? literature review.
\newblock {\em International Finance Corporation (IFC). Washington, DC}, 2002.

\bibitem{chawla2003smoteboost}
N.~Chawla, A.~Lazarevic, L.~Hall, and K.~Bowyer.
\newblock Smoteboost: Improving prediction of the minority class in boosting.
\newblock {\em Knowledge Discovery in Databases: PKDD 2003}, pages 107--119,
  2003.

\bibitem{deyoung2015risk}
R.~DeYoung, A.~Gron, G.~Torna, and A.~Winton.
\newblock Risk overhang and loan portfolio decisions: small business loan
  supply before and during the financial crisis.
\newblock {\em The Journal of Finance}, 70(6):2451--2488, 2015.

\bibitem{garcia2010credit}
A.~Garcia-Tabuenca and J.~L. Crespo-Espert.
\newblock Credit guarantees and sme efficiency.
\newblock {\em Small Business Economics}, 35(1):113--128, 2010.

\bibitem{han2005borderline}
H.~Han, W.-Y. Wang, and B.-H. Mao.
\newblock Borderline-smote: a new over-sampling method in imbalanced data sets
  learning.
\newblock {\em Advances in intelligent computing}, pages 878--887, 2005.

\bibitem{hand1997statistical}
D.~J. Hand and W.~E. Henley.
\newblock Statistical classification methods in consumer credit scoring: a
  review.
\newblock {\em Journal of the Royal Statistical Society: Series A (Statistics
  in Society)}, 160(3):523--541, 1997.

\bibitem{he2009learning}
H.~He and E.~A. Garcia.
\newblock Learning from imbalanced data.
\newblock {\em IEEE Transactions on knowledge and data engineering},
  21(9):1263--1284, 2009.

\bibitem{irfan2011game}
M.~T. Irfan and L.~E. Ortiz.
\newblock A game-theoretic approach to influence in networks.
\newblock In {\em AAAI}, 2011.

\bibitem{jones2015empirical}
S.~Jones, D.~Johnstone, and R.~Wilson.
\newblock An empirical evaluation of the performance of binary classifiers in
  the prediction of credit ratings changes.
\newblock {\em Journal of Banking \& Finance}, 56:72--85, 2015.

\bibitem{khandani2010consumer}
A.~E. Khandani, A.~J. Kim, and A.~W. Lo.
\newblock Consumer credit-risk models via machine-learning algorithms.
\newblock {\em Journal of Banking \& Finance}, 34(11):2767--2787, 2010.

\bibitem{kovalerchuk2000data}
B.~Kovalerchuk and E.~Vityaev.
\newblock {\em Data mining in finance: advances in relational and hybrid
  methods}, volume 547.
\newblock Springer Science \& Business Media, 2000.

\bibitem{kubat1997addressing}
M.~Kubat, S.~Matwin, et~al.
\newblock Addressing the curse of imbalanced training sets: one-sided
  selection.
\newblock In {\em ICML}, volume~97, pages 179--186. Nashville, USA, 1997.

\bibitem{laurikkala2001improving}
J.~Laurikkala.
\newblock Improving identification of difficult small classes by balancing
  class distribution.
\newblock {\em Artificial Intelligence in Medicine}, pages 63--66, 2001.

\bibitem{levitsky1997credit}
J.~Levitsky.
\newblock Credit guarantee schemes for smes--an international review.
\newblock {\em Small Enterprise Development}, 8(2):4--17, 1997.

\bibitem{macleod1987re}
J.~E. Macleod, A.~Luk, and D.~M. Titterington.
\newblock A re-examination of the distance-weighted k-nearest neighbor
  classification rule.
\newblock {\em IEEE Transactions on Systems, Man, and Cybernetics},
  17(4):689--696, 1987.

\bibitem{meng2015credit}
X.~L.~X. Meng.
\newblock Credit risk evaluation for loan guarantee chain in china.
\newblock 2015.

\bibitem{niu2017hybrid}
Z.~Niu, D.~Cheng, J.~Yan, J.~Zhang, L.~Zhang, and H.~Zha.
\newblock A hybrid approach for risk assessment of loan guarantee network.
\newblock {\em arXiv preprint arXiv:1702.04642}, 2017.

\bibitem{dcheng2018pvis}
Z.~Niu, D.~Cheng, L.~Zhang, and J.~Zhang.
\newblock Visual analytics for network-guaranteed loan risk management.
\newblock In {\em Pacific Visualization Symposium (PacificVis), 2018 IEEE}.
  IEEE, 2018.

\bibitem{onnela2006complex}
J.-P. Onnela et~al.
\newblock {\em Complex networks in the study of financial and social systems}.
\newblock Helsinki University of Technology, 2006.

\bibitem{peterson2009k}
L.~E. Peterson.
\newblock K-nearest neighbor.
\newblock {\em Scholarpedia}, 4(2):1883, 2009.

\bibitem{ruzzier2006sme}
M.~Ruzzier, R.~D. Hisrich, and B.~Antoncic.
\newblock Sme internationalization research: past, present, and future.
\newblock {\em Journal of small business and enterprise development},
  13(4):476--497, 2006.

\bibitem{tang2009social}
J.~Tang, J.~Sun, C.~Wang, and Z.~Yang.
\newblock Social influence analysis in large-scale networks.
\newblock In {\em Proceedings of the 15th ACM SIGKDD international conference
  on Knowledge discovery and data mining}, pages 807--816. ACM, 2009.

\bibitem{wu2014business}
D.~D. Wu, S.-H. Chen, and D.~L. Olson.
\newblock Business intelligence in risk management: Some recent progresses.
\newblock {\em Information Sciences}, 256:1--7, 2014.

\bibitem{xiao2016modeling}
S.~Xiao, J.~Yan, C.~Li, B.~Jin, X.~Wang, X.~Yang, S.~M. Chu, and H.~Zha.
\newblock On modeling and predicting individual paper citation count over time.
\newblock In {\em IJCAI}, pages 2676--2682, 2016.

\bibitem{yan2015sales}
J.~Yan, M.~Gong, C.~Sun, J.~Huang, and S.~M. Chu.
\newblock Sales pipeline win propensity prediction: a regression approach.
\newblock In {\em Integrated Network Management (IM), 2015 IFIP/IEEE
  International Symposium on}, pages 854--857. IEEE, 2015.

\bibitem{yan2011incremental}
J.~Yan, C.~Tian, J.~Huang, and F.~Albertao.
\newblock Incremental dictionary learning for fault detection with applications
  to oil pipeline leakage detection.
\newblock {\em Electronics letters}, 47(21):1198--1199, 2011.

\bibitem{yu2011credit}
L.~Yu, X.~Yao, S.~Wang, and K.~K. Lai.
\newblock Credit risk evaluation using a weighted least squares svm classifier
  with design of experiment for parameter selection.
\newblock {\em Expert Systems with Applications}, 38(12):15392--15399, 2011.

\bibitem{zhou2006training}
Z.-H. Zhou and X.-Y. Liu.
\newblock Training cost-sensitive neural networks with methods addressing the
  class imbalance problem.
\newblock {\em IEEE Transactions on Knowledge and Data Engineering},
  18(1):63--77, 2006.

\end{thebibliography}

\end{document}